\documentclass[aps,pre,twocolumn,superscriptaddress,showpacs]{revtex4}
\usepackage{epsfig}
\usepackage{amsmath}
\usepackage{times}

\begin{document}

\title{An adaptive routing strategy for packet delivery in complex networks}

\author{Huan Zhang }
\affiliation{Institute of theoretical physics and Department of
Physics, East China Normal University, Shanghai, 200062, P. R.
China}
\author{Zonghua Liu }
\affiliation{Institute of theoretical physics and Department of
Physics, East China Normal University, Shanghai, 200062, P. R.
China}
\author{Ming Tang}
\affiliation{Institute of theoretical physics and Department of
Physics, East China Normal University, Shanghai, 200062, P. R.
China}
\author{P. M. Hui}
\affiliation{Department of Physics, The Chinese University of Hong
Kong,\\ Shatin, New Territories, Hong Kong}

\date{25 June}

\begin{abstract}
We present an efficient routing approach for delivering packets in
complex networks.  On delivering a message from a node to a
destination, a node forwards the message to a neighbor by
estimating the waiting time along the shortest path from each of
its neighbors to the destination.  This projected waiting time is
dynamical in nature and the path through which a message is
delivered would be adapted to the distribution of messages in the
network.  Implementing the approach on scale-free networks, we
show that the present approach performs better than the
shortest-path approach and another approach that takes into
account of the waiting time only at the neighboring nodes.  Key
features in numerical results are explained by a mean field
theory.  The approach has the merit that messages are distributed
among the nodes according to the capabilities of the nodes in
handling messages.
\end{abstract}
\pacs{89.75.Fb,89.20.-a,05.70.Jk} \maketitle

\section{Introduction}
The problem of traffic congestions in communication networks is
undoubtedly an important issue.  The problem is related to the
geometry of the underlying network, the rate that messages are
generated and delivered, and the routing strategy.  Many studies
have been focused on spatial structures such as regular lattices
and the Cayley tree
\cite{Li:1989,Taqqu:1997,Crove:1997,Falout:1999,Toru:1998,Fuks:1999,Arena:2001,
Guim:2002,Woolf:2002,Valv:2002}.  Random networks and scale-free
(SF) networks have also been widely studied.  The former is
homogeneous with a Poisson degree distribution; while the latter
typically exhibits a power-law degree distribution of the form
$P(k) \sim k^{-\gamma}$ signifying the existence of nodes with
large degrees. SF networks are found in many real-world networks,
such as the Internet, World Wide Web (WWW), and metabolic network
\cite{AB:2002,Mendes:2003,Boccaletti:2006}.  A standard model of
SF networks is the Barab\'{a}si and Albert (BA) model of growing
networks with preferential attachments \cite{BA:1999,BAJ:1999}.
The BA model gives a degree distribution of $P(k) \sim k^{-3}$ and
is non-assortative \cite{Newman:2002,Catan:2005}, i.e., the chance
of two nodes being connected is independent of the degrees of the
nodes concerned.  While there are other variations on the BA
models that give a degree exponent that deviates from $3$
\cite{Doro:2000,Krap:2000,Cald:2002,Liu:2002,Liu:2002a}, the BA
model still serves as the basic model for SF networks.  In the
present work, we study how a dynamical and adaptive routing
strategy would enhance the performance in delivering messages
over a BA scale-free network.

In communication models, the nodes are taken to be both hosts and
routers, and the links serve as possible pathways through which
messages or packets are forwarded to their destination. Early
studies assumed a constant (degree-independent) packet generation
rate $\lambda$ and a constant rate of delivering one packet per
time step at each node.  As $\lambda$ increases, the traffic goes
from a free-flow phase to a congested or jamming phase. Obviously,
such models are too simple for real-world networks. More realistic
models should incorporate the fact that nodes with higher degrees
would have higher capability of handling information packets and
at the same time generate more packets. Models with
degree-dependent packet-delivery rate of the form $(1 + \beta
k_{i})$ \cite{Zhao:2005,Liu:2006} and degree-dependent
packet-generating rate \cite{Liu:2006} of the form $\lambda k_{i}$
have recently been proposed and studied, with routing strategy
based on forwarding messages through the shortest path to their
destination.  Implementing this routing strategy in SF and random
networks indicate that it is easier to lead to congestions in SF
networks than in random networks \cite{Zhao:2005,Liu:2006}.  It is
because the nodes with large degrees in SF networks are on many
shortest-paths between two arbitrarily chosen nodes, i.e., large
betweenness \cite{Goh:2001}. Many packets will be passing by and
queueing up at these nodes enroute to their destination.  In a
random network, jamming is harder to occur as the packets tend to
be distributed quite uniformly to each node.

A good routing algorithm is essential for sustaining the proper
functioning of a network
\cite{Guimera:2002,Yan:2006,Moreno:2003,Adilson:2004}.  The
shortest-path routing approach is based on {\em static}
information, i.e., once the network is constructed, the
shortest-paths are fixed.  To improve routing efficiency,
Echenique {\em et al.} \cite{Echenique:2004,Echenique:2005}
proposed an approach in which a node would choose a neighboring
node to deliver a packet by considering the shortest-path from the
neighboring node to the destination {\em and} the waiting time at
the neighboring node. The waiting time depends on the number of
packets in the queue at a neighboring node at the time of decision
and thus corresponds to a {\em dynamical} or {\em time-dependent}
information. This algorithm performs better than the shortest-path
approach, as packets may be delivered not necessarily through the
shortest-path and thus the loading at the higher degree nodes in a
SF network is reduced. The approach has also been applied to
networks with degree-dependent packet generation rate
\cite{Chen:2006}.  Recently, Wang {\em et al.} \cite{Wang:2006}
proposed an algorithm that tends to spread the packets evenly to
nodes by considering information on nearest neighbors. However,
the delivering time turns out to be much longer than that in the
shortest-path approach as the packets tend to wander around the
network.

In the present work, we propose an efficient routing strategy that
is based on the projected waiting time along the shortest-path
from a neighboring node to the destination.  The algorithm is
implemented in BA scale-free networks, with degree-dependent
packet generating and delivering rates.  Results show that jamming
is harder to occur using the present strategy, when compared with
both the shortest-path approach and the Echenique's approach.  Key
features observed in numerical results are explained within a mean
field treatment.  The present approach has the advantage of
spreading the packets among the nodes according to the degrees of
the nodes. In this way, every node can contribute to the packet
delivery process.

The paper is organized as follows.  The model, including the
underlying network, the packet generation and delivery mechanisms,
and routing strategy, is introduced in Sec.II.  In Sec.III, we
present numerical results and compared them with those of the
other routing strategies.  We also explain key features within a
mean field theory.  We summarize our results in Sec.IV.

\section{Model}
The underlying network structure is taken to be the
Barabasi-Albert (BA) scale-free growing network with $N$ nodes
\cite{BAJ:1999}. Starting with $m_{0}$ nodes, each new node
entering the network is allowed to establish $m$ new links to
existing nodes. Preferential attachment whereby an existing node
$i$ with a higher degree $k_{i}$ has a higher probability $\Pi_{i}
\sim k_{i}$ to attract a new link is imposed.  The mean degree of
the network is $\langle k \rangle = 2m$ and the degree
distribution $P(k)$ follows a power-law behavior of the form $P(k)
\sim k^{-3}$.

The dynamics of packet generation and delivery is implemented as
follows.  Due to the inhomogeneous nature of the BA network, it is
more natural to impose a packet generation rate that is
proportional to the degree of a node. At each time step, a node
$i$ creates $\lambda k_{i}$ new packets.  The fractional part of
$\lambda k_{i}$ is implemented probabilistically.  A destination
is randomly assigned to each created packet.  The newly created
packets will be put in a queue at the node and delivery will be
made on the first-in-first-out basis.  The packets in the queue
may consist of those which are created at previous time steps and
received from neighboring nodes enroute to their destination. We
also assume a packet delivery rate that is proportional to the
degree of a node \cite{Liu:2006}. At each time step, a node $i$
delivers at most $(1+\beta k_{i})$ packets to its neighbors.  The
fractional part of $(1+\beta k_{i})$ is implemented
probabilistically.  A larger $\beta$ implies a higher
packet-handling capability, but it would translate into higher
cost or capital.  Here, the parameters $\lambda$ and $\beta$ are
taken to be node-independent.  A packet is removed from the system
upon arrival at its destination. For a given generation rate
characterized by $\lambda$, there exists a critical value of the
delivery rate $\beta_{c}$ such that for delivery rates $\beta <
\beta_{c}$, packets tend to accumulate in the network resulting in
a jamming phase; while for $\beta > \beta_{c}$, a non-jamming
phase results as there are as many packets delivered to their
destination as created.  A better performance is thus
characterized by a smaller value of $\beta_{c}$.

The novel feature of the present work is the routing strategy or
the selection of a neighbor in delivering a packet. The idea is to
choose a neighbor that would give the shortest time, including
waiting time, to deliver the packets along the shortest path from
the chosen neighbor to the destination. Consider a packet with
destination node $j$ leaving node $i$. Each of the $k_{i}$
neighbors of node $i$ has a shortest path to the destination node
$j$.  The shortest path refers to the smallest number of links
from a node to another. However, due to the possible accumulation
of packets at each node, the number of time steps it takes to
deliver the message may be different from the number of links
along the shortest path. Consider a neighbor labelled $\ell$ of
the node $i$.  We label the shortest path from node $\ell$ to $j$
by $\{SP:\ell,j\}$. Along this path, we evaluate the following
quantity for the node $\ell$:
\begin{equation}
d(\ell) = \sum_{s \in \{SP:\ell,j\}} \frac{n_{s}}{1 + \beta
k_{s}},
\end{equation}
where the sum is over the nodes along the shortest path
$\{SP:\ell,j\}$, excluding the destination.  Here, $n_{s}$ is the
number of packets accumulated at node $s$, at the moment of
decision. Thus, $d(\ell)$ is an estimate of the time that a
packet would take to go from node $\ell$ to the destination $j$
through the shortest path.  Node $i$ would choose a neighboring
node with the minimum $d(\ell)$ to forward the packet, i.e., the
selection is based on $min\{d(\ell), \ell \in \{i\}\}$, where
$\{i\}$ is the set of $k_{i}$ nodes consisting of the neighbors
of node $i$. This procedure is repeated for each node and each
packet in every time step. For a network far from jamming, each
node can handle all the packets in every time step.   In this
free-flow situation, the quantity $d(\ell)$ simply measures the
shortest path $d_{\ell,j}$ from $\ell$ to $j$.  When packets are
queueing up at the nodes, however, a delivery mechanism based on
$d(\ell)$ takes into account of the queueing time and may not
pass the packet to a neighboring node that is closest to the
destination.

To justify our routing scheme, we will compare results with two
other routing strategies widely studied in the literature. Using
the same packet generating mechanism, the shortest-path approach
selects a neighbor with the shortest path to the destination for
forwarding a packet.  Echenique {\em et al.}
\cite{Echenique:2004,Echenique:2005} proposed an approach that
takes into account of the waiting time.  For a delivering rate of
one packet per time step, they proposed to choose a neighbor that
has a minimum value of $h d_{\ell,j} + (1-h) n_{\ell}$, where
$d_{\ell,j}$ is the shortest path length from node $\ell$ to $j$.
The parameter $h$ is a weighing factor, which can be taken as a
variational parameter and $h \approx 0.8$ is found to give the
best performance.  The Echenique's approach thus accounts for the
waiting time only at the neighboring nodes.  For a delivery rate
of $(1+\beta k_{i})$, a modified Echenique's approach is to
choose a neighboring node with a minimum value of
\begin{equation}
\delta_{\ell} = h d_{\ell,j} + (1 - h)\frac{n_{\ell}}{1 + \beta
k_{\ell}}.
\end{equation}
We have checked that for a given value of $\lambda$, the smallest
value of $\beta_{c}$ is attained for values of $h \sim 0.8$ to
$0.85$.  In what follows, we will use a value of $h=0.8$ for the
Echenique's approach given by Eq.(2).

\section{Results and Discussion}
\begin{figure}
\epsfig{figure=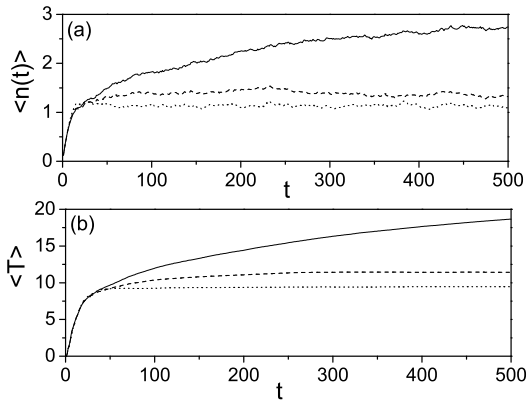,width=0.9\linewidth}\vspace{-0.5cm}
\caption{(a) Numerical results for the average number of packets
per node $<n(t)>$ and (b) the average delivering time $<T>$ as a
function of time for $\lambda=0.02$.  Lines from top to bottom
correspond to $\beta=0.04$, $0.048$ and $0.07$.} \label{evolution}
\end{figure}
The different phases in a network can be illustrated by looking at
the average number of packets per node at a given time $\langle
n(t) \rangle$ and the average time for a packet to remain in the
network or the delivering time $\langle T \rangle$.  We take
$m_{0}=3$ and $m=3$ and construct a BA scale-free network of
$N=1000$ nodes. Figure \ref{evolution} shows the results of
$\langle n(t) \rangle$ and $\langle T \rangle$ as a function of
time for a fixed value of $\lambda = 0.02$.  As $\beta$ increases,
there are distinct behavior. For values of $\beta$ smaller than
some critical value $\beta_{c}(\lambda)$, $\langle n(t) \rangle$
grows almost linearly with time after the transient (see
Fig.\ref{evolution}(a)). This corresponds to a jamming phase.  As
$\beta$ increases, the slope in the long time behavior decreases,
indicating a slower accumulation of packets in the network as the
ability of handling packets $\beta$ increases.  For $\beta >
\beta_c(\lambda)$, $\langle n(t) \rangle$ becomes independent of
time in the long time limit.  This corresponds to a non-jamming
phase.  Similarly behavior is exhibited in $\langle T \rangle$. In
the jamming phase, $\langle T \rangle$ increases with time
monotonically, due to the increasing waiting time in the queues at
intermediate nodes as a packet is forwarded to its destination.
Fewer packets are delivered to their destination than generated.
In the non-jamming phase, $\langle T \rangle$ becomes independent
of time in the long time limit.  In this regime, further
increasing $\beta$ will lead to smaller $\langle n(t) \rangle$ and
shorter $\langle T \rangle$ in the long time limit until these
quantities saturate.  This is possible since a non-jamming phase
corresponds either to the case in which all the packets at the
nodes are forwarded every time step or steady queues of packets
exist at the nodes.  In both cases, the number of packets does not
increase in the long time limit.  The former case is the free-flow
phase, while the latter is reminiscent of the synchronized phase
in vehicular traffic flows \cite{trafficpaper} in which the
packets undergo a stop-and-go behavior.  For $\beta = 0.07 >
\beta_{c}$ for example, $\langle T \rangle \approx 9.5$, which is
somewhat larger than the average shortest distance or diameter $D
\approx 3.332$ of the network. This indicates that, due to the
routing strategy in forwarding a packet, the dynamics in the
free-flow phase is different from that of the shortest-path
approach.

\begin{figure}
\epsfig{figure=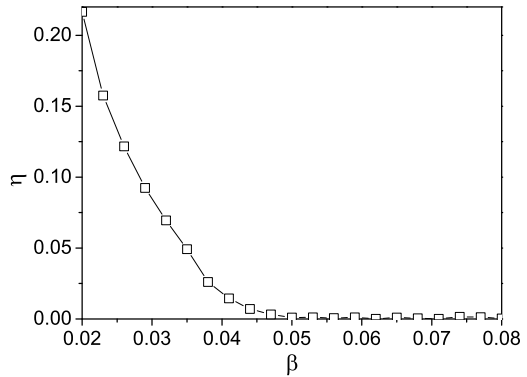,width=0.9\linewidth}\vspace{-0.5cm}
\caption{Numerical results for the quantity $\eta$ as function of
$\beta$ for $\lambda=0.02$.  The critical value $\beta_{c}$
separates the behavior of $\eta =0$ for $\beta > \beta_{c}$ and
$\eta \neq 0$ for $\beta < \beta_{c}$.} \label{order}
\end{figure}
\begin{figure}
\epsfig{figure=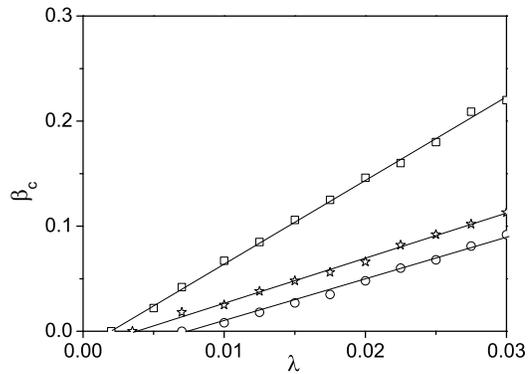,width=0.9\linewidth}\vspace{-0.5cm}
\caption{The critical value $\beta_c(\lambda)$ for three different
routing approaches for forwarding packets: the present approach
(circles), the Echenique's approach with $h=0.8$ (stars), and the
shortest-path approach (squares).  The lines are guides to eye.}
\label{betac}
\end{figure}

The critical value $\beta_c(\lambda)$ can be determined by
considering the quantity
\begin{equation}\label{eq:slope}
\eta = \lim_{t\rightarrow \infty}\frac{1}{2m\lambda}\frac{<\Delta
n>}{\Delta t},
\end{equation}
where $\Delta n=n(t+\Delta t)-n(t)$ and the average is over all
the nodes at a time $t$.  This quantity $\eta \in [0,1]$ is
basically the slope of $\langle n(t) \rangle$ in the long time
limit.  In the non-jamming phase, the slope vanishes and $\eta=0$;
while in the jamming phase, $\eta>0$.  Figure \ref{order} shows
$\eta$ as a function of $\beta$, for a fixed value of $\lambda =
0.02$. The critical value $\beta_{c}$ can be identified as the
value that separates the $\eta=0$ and $\eta \neq 0$ behavior.  We
carried out similar calculations for different values of
$\lambda$ and determined $\beta_{c}(\lambda)$.  The results are
shown in Fig.\ref{betac} (circles).  We will explain the form of
$\beta_{c}(\lambda)$ using a mean field theory.  The curve
$\beta_{c}(\lambda)$ can also be regarded as a phase boundary in
the $\lambda$-$\beta$ plane, separating the jamming phase below
the curve and the non-jamming phase above the curve.

To show the superior performance of our routing strategy, we also
performed calculations using the shortest-path approach and the
Echenique's approach with $h=0.8$ in Eq.(2).  The same
degree-dependent packet generating mechanism is used. Results of
$\beta_{c}(\lambda)$ for these two models are also shown in
Fig.\ref{betac} for comparison.  The present approach gives the
best performance.  For a given $\lambda$, we see the improvement
in performance from the shortest-path approach through the
Echenique's approach to the present approach, signified by the
drop of $\beta_{c}$. For the shortest-path approach, it has been
shown \cite{Liu:2006} that $\beta_{c}(\lambda)$ follows the
functional form of
\begin{equation}\label{eq:SP}
\beta_{c}^{SP} = \alpha D (\lambda - \lambda_{min}^{SP}),
\end{equation}
where $\alpha \approx 2$ and $\lambda_{min}^{SP} = 1/(\alpha D
k_{max})$ with $D$ being the diameter and $k_{max}$ the maximum
degree of the network.  For $\lambda < \lambda_{min}^{SP}$,
$\beta_{c}^{SP}=0$.  With the present approach,
$\beta_{c}(\lambda)$ follows a similar functional form, but with a
{\em higher} value of $\lambda_{min}$ and a smaller prefactor that
gives the slope.  Both the present approach and the Echenique's
approach perform better than the shortest-approach approach
because packets are re-directed to other nodes when there are
long queues at the hubs.

\begin{figure}
\epsfig{figure=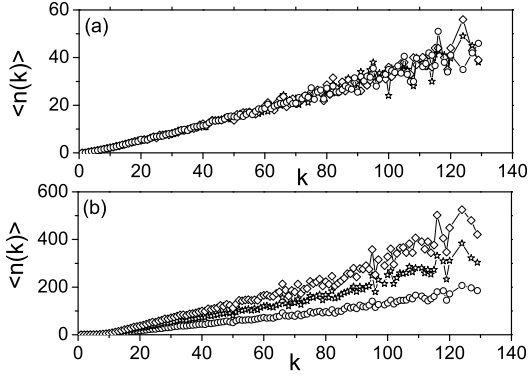,width=0.9\linewidth}\vspace{-0.5cm}
\caption{The average number of packets $n_{k}$ at a node of degree
$k$ as a function of $k$ for $\lambda=0.02$ for (a) $\beta = 0.06
(>\beta_{c})$ and (b) $\beta = 0.04 (< \beta_{c})$.  In each case,
results are shown at three different times of $t=100$ (circles),
$200$ (stars) and $300$ (squares) time steps.} \label{stationary}
\end{figure}
The better performance of the present approach is achieved by
spreading the packets among the nodes so that the number of
packets at a node is proportional to the degree $k$ of the node in
the free-flow phase.  We use a mean field approach to illustrate
this point.  Let $n_{k}$ be the average number of packets at the
nodes with degree $k$.  In the free-flow phase where $n_{k} < 1 +
\beta k$, we have
\begin{equation}\label{eq:smallnk}
\frac{dn_k(t)}{dt}= \lambda k - n_{k}(t) + k
\sum_{k'=k_{min}}^{k_{max}} P(k'|k)\frac{n_{k'}(t)}{k'} - \lambda
\langle k \rangle.
\end{equation}
The first and second terms denote the packets generated at the
node and delivered to neighboring nodes, respectively.  The third
term accounts for the packets delivered {\em into} the node from
its neighboring nodes.  Here $P(k'|k)$ is the conditional
probability that a node of degree $k$ has a neighbor of degree
$k'$ and the sum runs from the minimum degree $k_{min}$ to
$k_{max}$ in the network.  In the free-flow regime, the packets
that are removed upon arrival at their destination can be assumed
to be $k$-independent and approximated by the term $\lambda
\langle k \rangle$.  The non-assortative feature of BA networks
\cite{Newman:2002,Catan:2005} gives $P(k'|k) = k'P(k')/\langle k
\rangle$, where $P(k')$ is the degree distribution.  After the
transient behavior, $dn_{k}/dt =0$ and we have
\begin{equation}\label{eq:stationary}
n_k= (\lambda+\frac{<n>}{<k>}) k - \lambda \langle k \rangle,
\end{equation}
where $\langle n \rangle = \sum_{k'=k_{min}}^{k_{max}}
P(k')n_{k'}$ is the mean number of packets per node.  Thus for $k
> \langle k \rangle =6$, $n_{k} \sim k$ in the free-flow phase after the
transient. Figure \ref{stationary}(a) shows the numerical results
obtained by averaging the number of packets on the nodes with
degree $k$ at different times (time $t=100$, $200$, $300$ time
steps) of a run. In the free-flow phase, $n_{k} \sim k$ and
becomes time-independent after the transient, as shown in
Fig.\ref{stationary}(a) for the case of $\beta = 0.06$ and
$\lambda = 0.02$.  This behavior is consistent with that in
Eq.(\ref{eq:stationary}).

For the jamming phase, numerical results (see
Fig.\ref{stationary}(b)) show that (i) $n_{k} \sim k$ at a fixed
instant {\em and} (ii) $n_{k}$ increases with time for fixed value
of $k$.  This behavior can be understood provided that the packets
are still distributed among the nodes in proportion to the degree
$k$ of a node via our strategy.  In this phase, the long time
behavior is characterized by an increasing accumulation of
packets and the delivery to destinations becomes negligible
compared with packet generation.  With $n_{k}> 1 + \beta k$ for
all nodes and ignoring the removal of packets, Eq.(6) is modified
to
\begin{eqnarray}\label{eq:largenk}
\frac{dn_k(t)}{dt}&=&  \lambda k -(1+\beta k)
+k\sum_{k'=k_{min}}^{k_{max}}
P(k'|k)\frac{1+\beta k'}{k'}\nonumber \\
&=&-1+k(\lambda+\frac{1}{<k>}).
\end{eqnarray}
It follows that $n_{k}(t)$ increases with time $t$ as
\begin{equation}\label{eq:evolution}
n_k(t)=n_k(0)+(k(\lambda+\frac{1}{<k>})-1)t,
\end{equation}
which describes very well the features in Fig.\ref{stationary}(b).
Thus, the present approach has the effect of reducing (increasing)
the probability of passing packets to neighbors with high (low)
degrees when there are long (no or short) queues, resulting in a
distribution of packets according to the degrees of the nodes.

A rough estimate of $\beta_{c}(\lambda)$ can be obtained by
equating $n_{k}$ in the free-flow phase to $(1+\beta_{c} k)$.  In
particular, taking $n_{k_{max}}=1+\beta_c k_{max}$, we get from
Eq.(\ref{eq:stationary}) that
\begin{eqnarray}\label{eq:critical}
\beta_c&=&\lambda - \lambda \frac{\langle k \rangle}{k_{max}}
+\frac{<n>}{<k>}-\frac{1}{k_{max}}\nonumber \\
&=&\lambda - \lambda \frac{\langle k \rangle}{k_{max}}
+\frac{(D-1)\sum \lambda k_i}{N<k>}-\frac{1}{k_{max}}\nonumber \\
&=& \left(D - \frac{\langle k \rangle}{k_{max}}\right)
\left(\lambda-\frac{1}{(D - \langle k \rangle/k_{max})
k_{max}}\right) \nonumber \\ & \approx & D \left(\lambda -
\frac{1}{D k_{max}}\right),
\end{eqnarray}
where $D$ is the average number of nodes that a packet passes
through from its origin to the destination, which is the diameter
of the network in the free-flow phase.  The last line is valid for
$k_{max} \gg \langle k \rangle$.  Comparing with Eq.(\ref{eq:SP})
for the shortest-path approach, we note that $\lambda_{min} =
1/(Dk_{max}) > \lambda_{min}^{SP}$ and the prefactor $D$, which
gives the slope in Fig.\ref{betac}, is smaller than that in the
shortest-path approach.  These features are consistent with
numerical results.  In particular, for $N=1000$ nodes, we found
that $D \approx 3.332$ and $k_{max} \approx 85$, giving
$\lambda_{min} \approx 0.007$, which is in reasonable agreement
with numerical results in Fig.\ref{betac}.

\section{Summary}
In summary, we have proposed an efficient routing strategy on
forwarding packets in a scale-free network.  The strategy accounts
not only for the physical separation from the destination but also
on the waiting time along possible paths.  We showed that our
strategy performs better than both the shortest-path approach and
the Echenique's approach.  Analytically, we construct a mean field
treatment which gives results in agreement with observed features
in numerical results.  Our routing strategy has the merit of
distributing the packets among the nodes according to the degree,
and hence handling capability, of the nodes.  Although our
discussion was carried out on BA networks, we believe that our
approach is also applicable in other spatial structures.

We end by comparing the three different routing strategies in more
general terms.  The shortest-path approach depends entirely on
geometrical information that is {\em static}.  Once the origin and
the destination of a packet is known, the shortest-path is fixed.
This strategy is {\em non-adaptive}, i.e., it will not be change
with time.  The Echenique's approach considers both geometrical
and local dynamical information.  By considering the waiting time
at a neighboring node, a packet from a node $i$ to a destination
$j$ will not always follow the same path.  Thus, the Echenique's
approach is a strategy that is {\em adaptive}, i.e., a decision
based on the current situation.  The present strategy, like the
Echenique's approach, is also adaptive and makes use of {\em
global} information in which all the waiting times along a path
are taken into consideration.  We see that by allowing for
adaptive strategies and taking more information into
consideration, a better performance results.  This line of thought
is in accordance with that in complex adaptive systems \cite{book}
whereby active agents may adapt, interact, and learn from past
experience.  It should be, however, noted that it pays to be
better. The shortest-path approach does not require update of the
routing strategy.  The Echenique's approach and the present
approach require continuing update of the number of packets
accumulated at the nodes.  Such updating plays the role of a cost,
with the payoff being the better performance.  Practical
implementation would have to consider the balance between the cost
and the payoff.

This work was supported by the NNSF of China under Grant No.
10475027 and No. 10635040, by the PPS under Grant No. 05PJ14036,
and by SPS under Grant No. 05SG27. P.M.H. acknowledges the support
from the Research Grants Council of the Hong Kong SAR Government
under grant number CUHK-401005.

Email: zhliu@phy.ecnu.edu.cn Suggested Referees:

\end{document}